\documentclass[aps,prb,reprint,amsmath,amssymb,showpacs,superscriptaddress,twocolumn]{revtex4-1}

\usepackage[colorlinks,citecolor=blue,linkcolor=blue]{hyperref}
\usepackage{graphicx}
\usepackage{dcolumn}
\usepackage{bm}

\newcommand{\IEF}{Institut d'Electronique Fondamentale, UMR CNRS 8622, Universit\'e Paris-Sud, 91405 Orsay, France}
\newcommand{\LPS}{Laboratoire de Physique des Solides, UMR CNRS 8502, Universit{\'e} Paris-Sud, 91405 Orsay, France}
\newcommand{\LSPM}{LSPM (CNRS-UPR 3407), Universit{\'e} Paris 13, Sorbonne Paris Cit{\'e}, 99 avenue Jean-Baptiste Cl{\'e}ment, 93430 Villetaneuse, France}

\begin{document}

\title{Interfacial Dzyaloshinskii-Moriya interaction in perpendicularly-magnetized Pt/Co/AlO$_x$ ultrathin films measured by Brillouin light spectroscopy}

\author{Mohamed Belmeguenai}
\email{belmeguenai.mohamed@univ-paris13.fr}
\affiliation{\LSPM}
\author{Jean-Paul Adam}
\email{jean-paul.adam@u-psud.fr}
\affiliation{\IEF}
\author{Yves Roussign{\'e}}
\affiliation{\LSPM}
\author{Sylvain Eimer}
\author{Thibaut Devolder}
\author{Joo-Von Kim}
\affiliation{\IEF}
\author{Salim Mourad Cherif}
\author{Andrey Stashkevich}
\affiliation{\LSPM}
\author{Andr{\'e} Thiaville}
\email{andre.thiaville@u-psud.fr}
\affiliation{\LPS}

\date{\today}

\begin{abstract}
Spin waves in perpendicularly-magnetized Pt/Co/AlO$_x$/Pt ultrathin films with varying Co thicknesses (0.6-1.2~nm) have been studied with Brillouin light spectroscopy in the Damon-Eshbach geometry. The measurements reveal a pronounced nonreciprocal propagation, which increases with decreasing Co thicknesses. This nonreciprocity is attributed to an interfacial Dzyaloshinskii-Moriya interaction (DMI), which is significantly stronger than asymmetries resulting from surface anisotropies for such modes. Results are consistent with an interfacial DMI constant $D_s = -1.7 \pm 0.11$~pJ/m, which favors left-handed chiral spin structures. This suggests that such films below 1~nm in thickness should support novel chiral states like skyrmions. 
\end{abstract}

\maketitle
In the magnetism of ultrathin films, it has recently been recognized that an antisymmetric exchange known as the Dzyaloshinskii-Moriya interaction (DMI) plays an important role in technologically relevant sputtered polycrystalline films \cite{brataas_spintronics:_2013}. This interaction can appear in thin film ferromagnets in contact with a material possessing strong spin-orbit coupling and is of interfacial origin \cite{fert_magnetic_1990}, which becomes all the more important as the film thickness decreases. The DMI modifies the statics~\cite{dzyaloshinskii_theory_1965,heide_dzyaloshinskii-moriya_2008} and dynamics~\cite{thiaville_dynamics_2012} of domain walls, and also stabilizes chiral nanoscale bubbles known as skyrmions~\cite{fert_skyrmions_2013}. As a number of applications have been proposed based on domain walls, both for data storage \cite{parkin_magnetic_2008} and logic gates \cite{allwood_magnetic_2005} -- concepts also applicable with skyrmions \cite{fert_skyrmions_2013} --, it is important to quantify the DMI in ultrathin films with the goal of controlling it by tailoring material structures. To achieve this, a reliable and direct method to quantify DMI in these ultrathin films is desirable.

In systems of interest for spintronics applications, the DMI is not expected to be sufficiently large to overcome the exchange interaction such that the uniform ferromagnetic state is destabilized. In these materials, estimates of the DMI have largely been based on how the interaction modifies the properties of domain walls. Because of the underlying symmetry of the interfacial DMI, its effect can be observed in ultrathin materials with an easy anisotropy axis perpendicular to the film plane, where a transition from achiral Bloch walls (favored by dipolar interactions) to homochiral N\'{e}el walls  \cite{heide_dzyaloshinskii-moriya_2008,chen_novel_2013} occurs for sufficiently large DMI. Since the DMI acts on the domain walls as an effective in-plane chiral field $\mu_0 H_{\rm DMI} = \pi D / M_s \Delta$, where $D$ is the micromagnetic DMI constant and $\Delta$ the domain wall width parameter, estimates for $H_{\rm DMI}$ can be obtained by applying an external in-plane field so as to counterbalance this DMI field. This idea underpins several recent experimental and theoretical studies, but in each case the determination of $D$ is at best indirect and rests upon strong assumptions involving the domain wall dynamics at hand. In particular, this has involved estimating changes in the wall energy in the creep regime \cite{je_asymmetric_2013,hrabec_measuring_2014}, interpreting current-driven motion assisted by applied fields with a one-dimensional model \cite{emori_current-driven_2013,ryu_chiral_2014,torrejon_interface_2014}, applying a droplet model to describe thermally-driven edge nucleation \cite{pizzini_chirality-induced_2014}, and extending the one-dimensional model to describe tilts in the domain wall structure across rectangular tracks \cite{boulle_domain_2013}. Other techniques involving the imaging of domain walls \cite{chen_novel_2013} or measurements of their stray fields \cite{tetienne_nanoscale_2014} require stabilizing domain walls in nanostructures and can only give bounds on the value of $D$ based on whether transitions from Bloch to N\'{e}el wall profiles are seen.

\begin{figure}
\centering
\includegraphics[width=8.5cm]{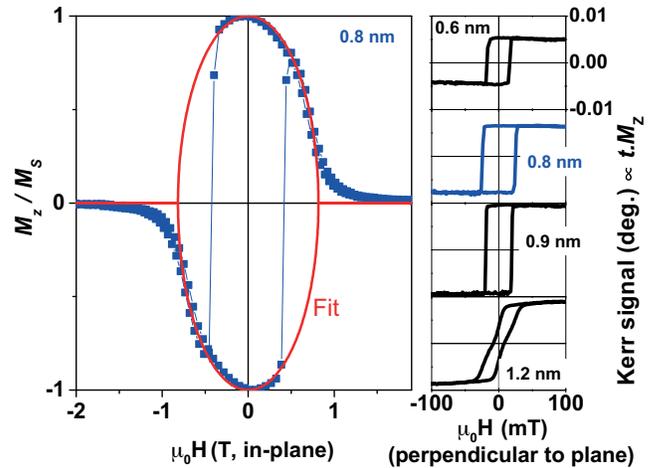}
\caption{(Color online) Polar MOKE magnetometry for Pt/Co($t$)/AlO$_x$ samples of different thicknesses $t$. Left panel: experimental (symbols) and fitted (red bold line) hysteresis loops of the perpendicular component of the magnetization versus in-plane applied magnetic field for the $t$=0.8~nm sample. Right panels: Hysteresis loops of some representative samples for fields applied perpendicularly to the sample plane.}
\label{fig:Magnetometry}
\end{figure}
\begin{figure}
\centering
\includegraphics[width=8.5 cm]{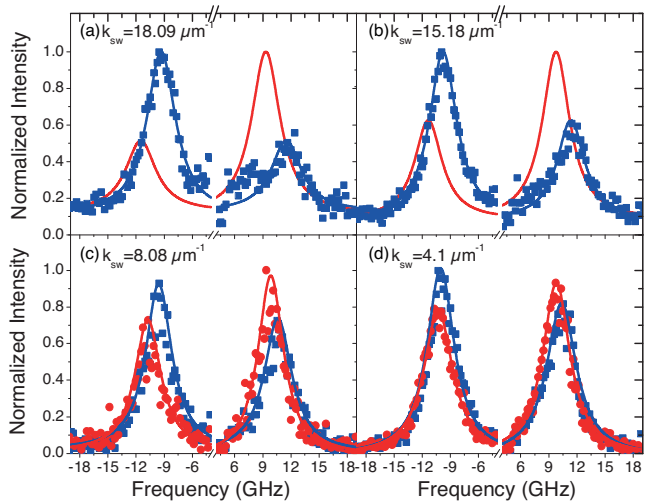}
\caption{ (Color online) BLS spectra measured for Pt/Co(1.2~nm)/AlO$_x$ at two different applied field values $\mu_0H$=0.4 T (resp. -0.4 T) in blue squares (resp. red circles) and at four characteristic light incidence angles corresponding to: $k_{sw}$= 18.09, 15.18, 8.08 and 4.1~$\mu$m$^{-1}$. Symbols refer to the experimental data and solid lines are the Lorentzian fits. For $k_\textrm{sw} = 18.09$ and 15.18~$\mu$m$^{-1}$, the curves for $\mu_0 H=-0.4$~T are those fitted for $\mu_0H=0.4$~T with inverted frequencies. }
\label{fig:BLSSpectra}
\end{figure}

Recently, a more direct measurement of $D$ through frequency shifts of oppositely propagating spin waves (SW) was proposed \cite{cortes-ortuno_influence_2013} and demonstrated experimentally for Pt/Co/Ni \cite{di_direct_2015}, Pt/NiFe \cite{nembach_spectroscopic_2014,stashkevich_non-reciprocity_2014} and Pt/CoFeB \cite{di_asymmetric_2015} films. Indeed, when magnetization and wave vector are both in-plane and perpendicular to each other, the chiral DMI interaction results in nonreciprocal SW propagation that manifests itself as a linear wave vector dependence of the SW frequencies. In this Letter, we provide a precise estimate using Brillouin light spectroscopy of the DMI in ultrathin Pt/Co/AlO$_x$ multilayers with perpendicular anisotropy, a system for which $D$ has been inferred to be very large based on \textit{ab initio} calculations \cite{freimuth_berry_2014} and previous experiments on domain walls \cite{pizzini_chirality-induced_2014, tetienne_nature_2014}. Furthermore, we show that the large Gilbert damping known for such systems is not an impediment for characterizing nonreciprocal SW propagation, since the frequency shifts are larger than the typical spectral line widths of the SW peaks.

Co ultrathin films with thickness varying from 0.6 to 1.2~nm were grown by sputtering in argon pressure of $2.5 \times 10^{-3}$~mTorr on Si/SiO$_2$ substrates buffered with a Ta(3 nm)/Pt(3 nm) bilayer, and then capped with AlO$_x$(2 nm)/Pt (3 nm) bilayer. In this system, the Pt bottom layer induces perpendicular magnetic anisotropy and DMI in the ultrathin Co layer, whereas the AlO$_x$ cap layer is thought to induce mainly perpendicular anisotropy. All experiments have been performed at room temperature. Magnetometry (AGFM and SQUID) has been used to measure hysteresis loops of the samples, with the field applied perpendicular to the sample plane, and to determine the magnetization at saturation, $M_\textrm{s}$. For the thicker samples, $M_\textrm{s}$ is similar to that of the bulk Co while that for the thinnest film is significantly smaller but remains in the range measured by Metaxas \emph{et al}.\cite{Metaxas07} (Tab.~\ref{tab:Parameters}). The  polar magneto-optical Kerr effect (p-MOKE) has been used to obtain the hysteresis loops with the magnetic field applied along the normal and in the plane of the sample (as shown in Fig.~\ref{fig:Magnetometry}). The Kerr rotation at saturation is deduced from the hysteresis loops obtained with the field applied perpendicular to the plane. This amplitude signal is then used to normalize the measured Kerr rotation while the field is applied in the sample plane (Fig.~\ref{fig:Magnetometry}). Effective anisotropy fields are then obtained from the hysteresis loops by fitting with the Stoner-Wohlfarth model. The obtained values are summarized in Table~\ref{tab:Parameters}. Figure~\ref{fig:Magnetometry} shows that all samples exhibit a perpendicular magnetic anisotropy that increases with decreasing Co thickness.

Brillouin light spectroscopy (BLS) gives access to SW modes with nonzero wave vectors. The SW in the film inelastically scatter the light from an incident laser beam. The frequency shift is analyzed using a 2$\times$3 pass Fabry-Perot interferometer, which typically gives access to a 3--300 GHz spectral frequency range. For the used backscattering study, the investigated spin wave vector lies in the plane of incidence and its length is $k_\textrm{sw} = 4\pi \sin(\theta_\textrm{inc})/\lambda$ (with $\theta_\textrm{inc}$ the angle of incidence and $\lambda=532$~nm the wavelength of the illuminating laser). Therefore, it can be swept in the $0-20 \mu$m$^{-1}$ interval through the rotation of the sample around a planar axis. The magnetic field was applied perpendicular to the incidence plane, which allows spin waves propagating along the in-plane direction perpendicular to the applied field to be probed (Damon-Eshbach (DE) geometry). 
For each angle of incidence, the spectra were obtained after counting photons up to 15 hours (especially for the highest incidence angles) to have well defined spectra where the line position can be determined with accuracy better than 0.1~GHz. The Stokes (S, negative frequency relative to the incident light as a SW was absorbed) and anti-Stokes (AS, positive frequency relative to the incident light as a SW was created) frequencies were then determined from Lorentzian fits to the BLS spectra. In the following, as we refer to the properties of the SW, $f_\mathrm{S}$ denotes the absolute value of the Stokes frequency, and wavevectors along that of the photons are counted positive.

The BLS measurements were performed with the magnetization saturated in the film plane under magnetic fields above the saturation fields deduced from the MOKE loops. Figure~\ref{fig:BLSSpectra} shows typical BLS spectra for the 1.2~nm thick sample for $k_\mathrm{sw}$=~18.09, 15.18, 8.08 and 4.1~$\mu m^{-1}$ corresponding to incidence angles $\theta_\textrm{inc}$=50, 40, 20 and 10$^\circ$, under an applied field $\mu_\textrm{0}H$=0.4~T. Importantly, mirror-symmetrical results were obtained for $\mu_\textrm{0}H$=-0.4~T, as expected from non-reciprocity. Beside the well-known intensity asymmetry of the S and AS modes due to the coupling mechanism between the light and SWs (in thin films), an unusually pronounced difference between the frequencies of both modes (non-reciprocity), especially for higher values of $k_\mathrm{sw}$, is revealed by these spectra. Various mechanisms, in particular perpendicular uniaxial surface anisotropy and DMI, can induce this frequency difference between the DE Stokes and anti-Stokes lines. However, an effect of interface anisotropy is observable only if the characteristic DE spatial asymmetry (of the dynamic magnetization distribution across the film) is sufficiently pronounced, in other words in relatively thick films such that $k_\mathrm{sw} t$ is not much smaller than unity. The frequency difference present in our samples is, despite the large interface anisotropy of Pt/Co, much larger than what is expected from different surface anisotropies at the two interfaces of the ferromagnetic film~\cite{stashkevich_non-reciprocity_2014}. We note also that the DMI effects seen are much larger than what was measured on perpendicularly magnetized Pt/CoFeB \cite{di_asymmetric_2015} and in-plane magnetized Pt/Co/Ni \cite{di_direct_2015} ultrathin films.

\begin{figure}[t]
\centering\includegraphics[width=6.5 cm]{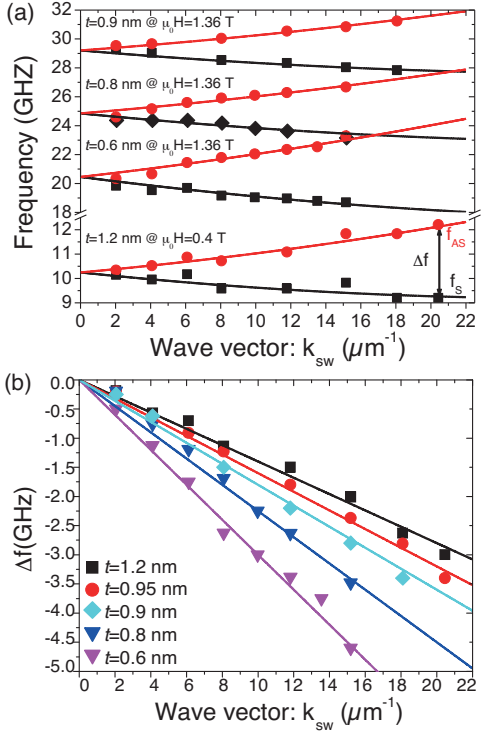}
\caption{(Color online) (a) Measured spin wave dispersion for various Co thicknesses $t$. Symbols show the experimental BLS data for AS (red, with $k_\mathrm{sw}$ inverted) and S frequencies (black). Solid lines represent the model described by Eq.~(1) with effective DMI constants $D_\textrm{eff}$ and magnetic parameters given in Tab.~\ref{tab:Parameters}. (b) Wave vector ($k_\textrm{sw}$) dependence of experimental frequency difference $\Delta f$ (symbols) compared to the DMI model (solid curves).}
\label{fig:curved}
\end{figure}

The variation of the frequencies of the S and AS modes as a function of the spin-wave wave vector is shown in Fig.~\ref{fig:curved}a (for the sake of clarity, the data for the $t$=0.95~nm sample are not presented). The prominent feature of these dispersion curves is their asymmetry with respect to $ k_\mathrm{sw} =0 $. 
The frequency difference $\Delta f = f_\mathrm{S} - f_\mathrm{AS}$ is plotted in Fig.~\ref{fig:curved}b as a function of $k_\mathrm{sw}$, revealing a linear dependence with a slope that changes markedly with Co thickness. For the samples studied here and for positive field, the AS mode frequency $f_\textrm{AS}$ was found to be always higher than $f_\textrm{S}$, as shown on Fig.~\ref{fig:curved}. It is worth noting that BLS measurements of $\Delta f$ as a function of $H$ (not shown) revealed that $\Delta f$ is independent of the applied field, as expected from the model mentioned below and similarly to previous work \cite{di_direct_2015}.

The BLS data have been analyzed using an analytical model relevant for ultrathin films, \cite{Moon13,Kostylev14,di_direct_2015} where the DE mode frequencies are given by
\begin{widetext}
\begin{equation}
\label{eq:f_sw}
f  = f_0 \pm\ f_\textrm{DMI}
   \equiv \frac{\gamma \mu_{0}} {2\pi} 
		\sqrt{ 
			\left(H+Jk_\textrm{sw}^2 + P(k_\textrm{sw}t)M_\textrm{s}\right)
			\left(H+Jk_\textrm{sw}^2 - P(k_\textrm{sw}t)M_\textrm{s} - H_{\rm K_{\rm eff}}\right)
} \pm\frac{\gamma}{\pi M_\textrm{s}} D_\textrm{eff}  k_\textrm{sw} .
\end{equation}
\end{widetext}
Here, $H$ represents the in-plane applied field, $M_\textrm{s}$ the saturation magnetization of Co, $\gamma$ the absolute value of the gyromagnetic ratio ($\gamma / (2 \pi)= g \times 13.996$~GHz/T, with $g$ the Land\'{e} factor), $\mu_0$ the vacuum permeability, $J = \frac{2A}{\mu_0 M_s}$ the SW stiffness (also called $D$ in the SW literature) with $A$ the micromagnetic exchange constant, $D_\textrm{eff}$ the effective micromagnetic DMI constant, $H_\textrm{K}$ the perpendicular uniaxial anisotropy field, $H_{\rm K_{\rm eff}} = H_\textrm{K} - M_\textrm{s}$ the effective anisotropy and $P(k_\textrm{sw}t) = 1 - \frac{1-\exp{\left(-\vert k_\textrm{sw}\vert t\right)}}{\vert k_\textrm{sw}\vert t}$.
In Eq.~(\ref{eq:f_sw}), the signs of $D$ and $k_\mathrm{sw}$ have been kept, $H$ is the absolute  field and $\pm$ its sign, according to the convention detailed below. 
If DMI is of purely interfacial origin, one expects a variation with thickness according to $D_\textrm{eff}(t)=D_s / t$.
From this, the frequency difference can be deduced to be
\begin{equation}
\Delta f = f_\textrm{S} - f_\textrm{AS} = 
\frac{2\gamma}{\pi M_s} D_\textrm{eff}k_\textrm{sw}
= \frac{2\gamma}{\pi}  k_\textrm{sw} \frac{D_\textrm{s}}{M_\mathrm{s}t},
\end{equation}
where the last equality stresses that in fact $D_s$ is the directly determined quantity, which is independent of the uncertainties in the film thickness $t$.
The experimental data were fitted conjointly by Eqs.~(1) and (2) by using the values of $M_s t$ determined from magnetometry and the bulk value $g=2.17$ to determine the different parameters summarized in Table~\ref{tab:Parameters}. 
We note the strength of the DMI found is largely insensitive to whether the exchange constant $A$ is assumed or fitted from the data, as different deformations of the dispersion curves are controlled by these parameters. 
For $t=0.6$~nm, $D_\textrm{eff}$ is in good agreement with the value $D=-2.2$~mJ/m$^2$ obtained by Ref.~\onlinecite{pizzini_chirality-induced_2014} from domain wall experiments.
The variations of $D_\mathrm{eff}$ with $t$ are consistent, to 10\% accuracy, with a single value for $D_s$.
It is also worth mentioning that the values of the saturation field in the sample plane, deduced from fits of the BLS data, are in very good agreement with those deduced from the MOKE hysteresis loops, (compare $H_\mathrm{sat}$ and $H_\mathrm{K_\mathrm{eff}}$ in Tab.~\ref{tab:Parameters}). This shows that any second-order anisotropy term can be neglected, and that the assumed value of $g$ is consistent with our findings.

Finally, we stress that the \emph{sign} of the DMI can be also determined. Indeed, performing the calculation leading to Eq.~(\ref{eq:f_sw}) shows that, when the transferred optical wave vector ($x$), the applied field ($y$) and the film normal from Pt to Co ($z$) form a right-handed reference frame, the Stokes SW corresponds to a left-handed cycloid. As we measured in this configuration and found that
$f_{AS} > f_S$, this means that DMI is negative, i.e., left-handed cycloids are favored.

\begin{table}
	\begin{center}
		\begin{tabular}{c||c|c|c||c|c}
$t$ & $\mu_0 M_\textrm{s}$ & $\mu_0 H_{\rm sat}$ & $\mu_0 H_{\rm K_{\rm eff}}$ & $D_\textrm{eff}$  & $D_s$\\
 & & MOKE & BLS & & \\
(nm) & (T) & (T) & (T) & (mJ/m$^2$) & (pJ/m)\\
\hline
\hline
0.6  & 1.38 & 0.95 & 1.03 & $2.71 \pm 0.16$ & $1.63 \pm 0.1$\\
0.8  & 1.48 & 0.82 & 0.87 & $2.18 \pm 0.25$ & $1.75 \pm 0.2$\\
0.9  & 1.51 & 0.75 & 0.68 & $1.88 \pm 0.08$ & $1.69 \pm 0.07$\\
0.95 & 1.68 & 0.51 & 0.36 & $1.76 \pm 0.24$ & $1.67 \pm 0.23$\\
1.2  & 1.71 & 0.10 & 0.11 & $1.57 \pm 0.18$ & $1.88 \pm 0.22$
		\end{tabular}
	\end{center}
    \caption{Magnetic parameters obtained from the best fits of BLS results with the above-mentioned model, using the saturation magnetization from magnetometry and $g=2.17$. The absolute values of $D_\mathrm{eff}$ and $D_s$ are given.}
    \label{tab:Parameters}
\end{table}
In conclusion, Pt/Co/AlO$_x$/Pt ultrathin films with perpendicular magnetization comprising various Co thicknesses have been studied by Brillouin light spectroscopy and magnetometry. The analysis of the BLS spectral reveals that the observed large nonreciprocities in the spin wave dispersion are consistent with a single surface DMI constant of $D_s=-1.7 \pm 0.11$~pJ/m. The sign of DMI implies that left-handed cycloidal spin structures are favored.
\begin{acknowledgments}
The authors would like to thank M. Kostylev and P. Moch for fruitful discussions, P. Balestri{\`e}re for help with the sample preparation, and J.-L. Cercus and L. Fruchter for help in SQUID measurements. This work has been partially supported by the Agence nationale de la recherche through contracts ANR-11-BS10-03 (NanoSWITI) and ANR-14-CE26-0012 (ULTRASKY), and by the R{\'{e}}gion {\^{I}}le-de-France through the DIM C'Nano (IMADYN project).
\end{acknowledgments}


\end{document}